\documentclass[10pt,letterpaper]{article}
\usepackage{opex3}
\usepackage{amssymb, amsfonts, amsmath}

\begin{document}

\title{Resonant absorption and amplification of circularly-polarized waves in inhomogeneous chiral media}

\author{Seulong Kim and Kihong Kim$^*$}
\address{Department of Energy Systems
Research and Department of Physics, Ajou University, Suwon 16499, Korea}

\email{$^*$khkim@ajou.ac.kr}

\begin{abstract}
It has been found that in the media where the dielectric permittivity $\epsilon$ or the magnetic permeability $\mu$ is near zero and in transition metamaterials where $\epsilon$ or $\mu$ changes from positive to negative values, there occur a strong absorption or amplification of the electromagnetic wave energy in the presence of an infinitesimally small damping or gain and a strong enhancement of the electromagnetic fields. We attribute these phenomena to the mode conversion of transverse electromagnetic waves into longitudinal plasma oscillations and its inverse process. In this paper, we study analogous phenomena occurring in chiral media theoretically using the invariant imbedding method. In uniform isotropic chiral media, right-circularly-polarized and left-circularly-polarized waves are the eigenmodes of propagation with different effective refractive indices $n_+$ and $n_-$, whereas in the chiral media with a nonuniform impedance variation, they are no longer the eigenmodes and are coupled to each other. We find that both in uniform chiral slabs where either $n_+$ or $n_-$ is near zero and in chiral transition metamaterials where $n_+$ or $n_-$ changes from positive to negative values, a strong absorption or amplification of circularly-polarized waves occurs in the presence of an infinitesimally small damping or gain. We present detailed calculations of the mode conversion coefficient, which measures the fraction of the electromagnetic wave energy absorbed into the medium, for various configurations of $\epsilon$ and $\mu$ with an emphasis on the influence of a nonuniform impedance. We propose possible applications of these phenomena to linear and nonlinear optical devices that react selectively to the helicity of the circular polarization.
\end{abstract}



\section{Introduction}

Because of the possibility of many exotic photonic phenomena,
there has been much recent interest in metamaterials with almost
zero dielectric permittivity $\epsilon$ and/or magnetic permeability $\mu$ [1--14]. These metamaterials have been termed epsilon-near-zero (ENZ), mu-near-zero (MNZ), or epsilon-and-mu-near-zero (EMNZ) metamaterials.
Another area which has drawn much research interest is
that of transition metamaterials, in which either $\epsilon$ or $\mu$ or both change continuously
from positive to negative values [15--21]. The common feature in both cases is the existence of the region where $\epsilon$, $\mu$, or both vanish
and the strong enhancement of the electromagnetic (EM) fields there. In the presence of an infinitesimally small amount of damping or gain in the resonance region, a large absorption or amplification of the EM wave energy occurs. The strong enhancement of the EM fields can also lead to enhanced nonlinear optical properties such as second harmonic generation and optical bistability \cite{4,6,12,20}.

In Ref.~15, it has been pointed out that
the strong absorption of the EM wave energy in transition metamaterials with vanishingly small damping is due to the mode conversion of transverse EM waves into longitudinal plasma oscillations.
The derivation of the fact that EM waves
in ordinary isotropic media are transverse requires that
$\epsilon$ and $\mu$ do not vanish anywhere. Therefore, in inhomogeneous media where
$\epsilon$ or $\mu$ vanishes in some region of space,
longitudinal EM waves can be excited.
A representative example is the mode conversion of EM waves
into longitudinal plasma waves at the resonance points where $\epsilon$
vanishes in cold unmagnetized plasmas [22--26]. The transfer of EM wave energy to the resonance region associated with more general kinds of mode
conversion occurring in magnetized plasmas plays a central role in a wide
range of phenomena in plasma physics [27--32].

In this paper, we are interested in a novel kind of
mode conversion occurring in isotropic chiral media.
These media are characterized by a coupling between the electric and magnetic fields, the strength of which
is denoted by a parameter $\gamma$ termed chiral index \cite{33,34}.
In the uniform case, right-circularly-polarized (RCP)
and left-circularly-polarized (LCP) waves are the eigenmodes of propagation. The effective refractive indices associated with RCP and LCP waves are $n_+$ ($=n+\gamma$) and $n_-$ ($=n-\gamma$) respectively, where $n$ is the ordinary refractive index. The difference between them gives rise to circular birefringence phenomena. In isotropic chiral media with a non-uniform impedance distribution, RCP and LCP waves are no longer eigenmodes and are coupled to each other.
In naturally occurring chiral media, the value of $\gamma$ is typically very small.
Recent rapid developments in metamaterials have made it possible to fabricate artificial chiral
metamaterials, which show a greatly enhanced effect of chirality.
Many interesting phenomena, including giant optical activity,
strong circular dichroism and chirality-induced negative refractive index,
can occur in these metamaterials [35--44].

As direct analogies to ENZ, MNZ, EMNZ and transition metamaterials, we consider uniform chiral slabs with near-zero $n_+$ or $n_-$ and chiral transition metamaterials, in which either $n_+$ or $n_-$ or both change from positive to negative values. We expect that the mode conversion of RCP or LCP waves or both into longitudinal modes and its inverse process occurs in these systems. In the presence of a vanishingly small amount of damping or gain in the resonance region, a large absorption or amplification of {\it circularly-polarized} waves will occur. When the impedance is not uniform throughout the space, there occurs a complication due to the coupling between RCP and LCP waves. This interesting mode conversion phenomenon has never been studied before.
We aim to explore this phenomenon by calculating
the mode conversion coefficient, which measures the wave absorption or amplification,
in a numerically exact manner using the invariant imbedding method [23, 31, 45--48].

In Sec.~\ref{sec2}, we describe the invariant imbedding method used in this paper. In Sec.~\ref{sec3}, we present the results of numerical calculations for uniform chiral slabs and chiral transition metamaterials. Finally, in Sec.~\ref{sec4}, we give a summary of the paper.

\section{Method}
\label{sec2}

In this section, we give a brief summary of the theoretical method used in the present study.
More details can be found in \cite{47}.
We consider isotropic chiral media, the constitutive relations of which are
given by
\begin{eqnarray}
\textbf{\textsl{D}}=\epsilon \textbf{\textsl{E}}+i\gamma
\textbf{\textsl{H}},~~ \textbf{\textsl{B}}=\mu
\textbf{\textsl{\textsl{H}}} - i\gamma \textbf{E}.
\end{eqnarray}
The scalar quantities $\epsilon$, $\mu$ and $\gamma$ are the dielectric
permittivity, the magnetic permeability and the chiral index
respectively \cite{34}.
We assume that the medium is stratified along the $z$ direction and $\epsilon$, $\mu$ and $\gamma$ are
functions of $z$ only. The wave is assumed to propagate in the $xz$ plane.
Then, using Eq.~(1) and the Maxwell's equations, we derive the coupled wave equations satisfied by the $y$ components of
the electric and magnetic fields, $E_y=E_y(z)$ and $H_y=H_y(z)$:
\begin{equation}
 \frac{{\rm d}^2 \psi}{{\rm d} z^2}
   - \frac{{\rm d} \mathcal{E}}{{\rm d} z}\mathcal{E}^{-1}(z)\frac{{\rm d} \psi}{{\rm d} z}
      + [k_0^2\mathcal{E}(z)\mathcal{M}(z)-q^2I]\psi = 0, \label{eq:psi}
\end{equation}
where $\psi=(E_y,H_y)^T$, $I$ is a $2\times 2$ unit matrix, $k_0$ ($=\omega/c$) is the vacuum wave number, $q$ is the $x$ component of the wave vector, and the $2\times 2$ matrix
functions $\mathcal{E}$ and $\mathcal{M}$ are given by
\begin{eqnarray}
\mathcal{E}=\left(\begin{array}{cc}\mu & i\gamma \\-i\gamma
& \epsilon
\\\end{array}\right),~~~\mathcal{M}=\left(\begin{array}{cc}\epsilon & i\gamma
\\-i\gamma & \mu\\\end{array}\right).\label{eq:epmu}
\end{eqnarray}

We assume that an inhomogeneous chiral medium
of thickness $L$ lies in the region $0\leq z\leq L$.
The waves are incident from the
vacuum region where $z>L$ and transmitted to another vacuum region
where $z<0$.
Our main interest is in calculating the $2\times 2$ matrix
reflection and transmission coefficients $r=r(L)$ and
$t=t(L)$, which we consider as functions of $L$.
The exact differential
equations satisfied by $r$ and $t$ have been derived from Eq.~(\ref{eq:psi}) previously,
using the invariant imbedding method \cite{47}:
\begin{eqnarray}
\frac{1}{ik_0\cos\theta}\frac{{\rm d}r}{{\rm d}l}&=&r\mathcal{E}+\mathcal{E}r
+\frac{1}{2}\left(r+I\right)\left[\mathcal{M}-\mathcal{E}+\tan^2\theta\left(\mathcal{M}
-\mathcal{E}^{-1}\right)\right] \left(r+I\right),
\label{eq:r}\\
\frac{1}{ik_0\cos\theta}\frac{{\rm d}t}{{\rm d}l}&=&t\mathcal{E}
+\frac{1}{2}t\left[\mathcal{M}-\mathcal{E}+\tan^2\theta\left(\mathcal{M}
-\mathcal{E}^{-1}\right)\right]\left(r+I\right),
\label{eq:t}
\end{eqnarray}
where $\theta$ is the incident angle.
We integrate the coupled differential equations (\ref{eq:r}) and (\ref{eq:t})
numerically from $l=0$ to $l=L$ using the initial conditions $r(0)=0$ and $t(0)=I$ and
obtain $r$ and
$t$ as functions of $L$.

The matrix component $r_{11}$ ($r_{21}$) is the reflection coefficient
when the incident wave is $s$-polarized and the transmitted wave is
$s$-polarized ($p$-polarized). Similar definitions are applied
to $r_{22}$ and $r_{12}$ and to the transmission coefficients. For circularly-polarized incident waves,
we obtain a new set of the reflection and transmission
coefficients $r_{ij}$ and $t_{ij}$, where $i$ and $j$
are either $+$ or $-$, from
\begin{eqnarray}
&&r_{++}=\frac{1}{2}(r_{22}+r_{11})+\frac{i}{2}(r_{21}-r_{12}),~~
r_{-+}=\frac{1}{2}(r_{22}-r_{11})+\frac{i}{2}(r_{21}+r_{12}),\nonumber\\
&&r_{+-}=\frac{1}{2}(r_{22}-r_{11})-\frac{i}{2}(r_{21}+r_{12}),~~
r_{--}=\frac{1}{2}(r_{22}+r_{11})-\frac{i}{2}(r_{21}-r_{12}),\nonumber\\
&&t_{++}=\frac{1}{2}(t_{22}+t_{11})+\frac{i}{2}(t_{21}-t_{12}),~~
t_{-+}=\frac{1}{2}(t_{22}-t_{11})+\frac{i}{2}(t_{21}+t_{12}),\nonumber\\
&&t_{+-}=\frac{1}{2}(t_{22}-t_{11})-\frac{i}{2}(t_{21}+t_{12}),~~
t_{--}=\frac{1}{2}(t_{22}+t_{11})-\frac{i}{2}(t_{21}-t_{12}).\label{eq:cir}
\end{eqnarray}
$r_{++}$ ($r_{-+}$) represents the
reflection coefficient when the incident wave is
RCP and the reflected wave is
RCP (LCP). $r_{--}$, $r_{+-}$ and $t_{ij}$'s are defined similarly.
The absorptance $A_1$ ($A_2$) is defined as the fraction of the incident wave energy absorbed into the medium when an $s$ ($p$) wave is incident. Similarly, $A_+$ ($A_-$) is the fraction of the incident wave energy absorbed into the medium when an RCP (LCP) wave
is incident.
These quantities can be obtained from
\begin{eqnarray}
A_1&=&1-\vert r_{11}\vert^2-\vert r_{21}\vert^2-\vert t_{11}\vert^2-\vert t_{21}\vert^2,\nonumber\\
A_2&=&1-\vert r_{12}\vert^2-\vert r_{22}\vert^2-\vert t_{12}\vert^2-\vert t_{22}\vert^2,\nonumber\\
A_+&=&1-\vert r_{++}\vert^2-\vert r_{-+}\vert^2-\vert t_{++}\vert^2-\vert t_{-+}\vert^2,\nonumber\\
A_-&=&1-\vert r_{+-}\vert^2-\vert r_{--}\vert^2-\vert t_{+-}\vert^2-\vert t_{--}\vert^2.
\end{eqnarray}
If $A_i$ ($i=1,2,+,-$) is negative, the wave is amplified inside the medium due to some gain mechanism.
In that case, $\vert A_i\vert$ ($=-A_i$) is a natural measure of wave amplification.
When mode conversion or its inverse process occurs, $A_i$ is nonzero even in the presence of
an infinitesimally small loss or gain. We define $\vert A_i\vert$ to be the mode conversion coefficient
in such cases.

When we deal with circularly-polarized waves propagating in weakly nonuniform chiral media,
it is convenient to introduce the fields ${\bf F}_{+}$ and ${\bf F}_{-}$
defined by
\begin{equation}
{\bf F_{\pm}}={\bf E}\pm i\eta{\bf H},~~~
\eta=\sqrt{\frac{\mu}{\epsilon}},
\end{equation}
where $\bf F_{+}$ and $\bf F_{-}$ describe RCP and LCP waves respectively and $\eta$ is the wave impedance.
These fields satisfy
\begin{equation}
\nabla\times{\bf F_{\pm}}=\pm\left[k_{\pm}{\bf F_{\pm}}+\frac{1}{2\eta}\nabla\eta\times\left({\bf F_{+}}-{\bf F_{-}}\right)\right],
~~~ k_{\pm}=n_{\pm}k_0,
\end{equation}
where
\begin{equation}
n_\pm=n\pm\gamma.
\end{equation}
In positive index media, the refractive index $n$ is equal to $\sqrt{\epsilon\mu}$, while in negative index media with simultaneously negative
real parts of $\epsilon$ and $\mu$, $n$ is equal to $-\sqrt{\epsilon\mu}$.
We notice that in uniform chiral media, RCP and
LCP waves are the eigenmodes of the wave equation with the
effective refractive indices $n_+$ and
$n_-$ respectively.
In inhomogeneous media with nonuniform impedance, these two modes are no longer eigenmodes and are coupled to each other .

\section{Numerical results}
\label{sec3}

\begin{figure}
\centering\includegraphics[width=8cm]{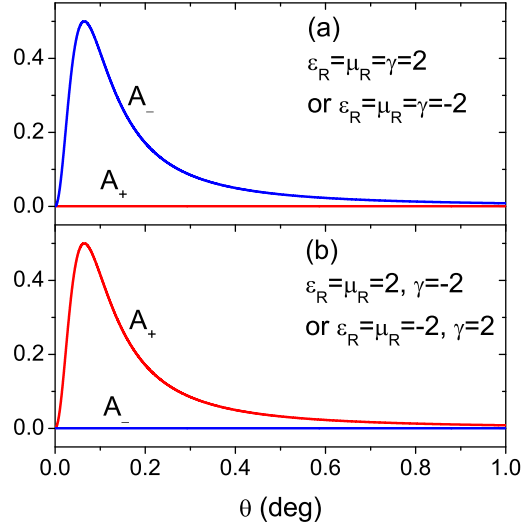}
\caption{Mode conversion coefficients $A_+$ and $A_-$ for RCP and LCP waves incident on a uniform chiral slab of thickness $L$ versus incident angle, when (a) $\epsilon_R=\mu_R=\gamma=\pm 2$ and (b) $\epsilon_R=\mu_R=-\gamma=\pm 2$. The other parameters used are $k_0L=5\pi$ and $\epsilon_I=\mu_I=10^{-5}$. In (a), strong absorption
for LCP waves occurs, while $A_+$ for RCP waves vanishes, because $n-\gamma=0$. In (b), strong absorption
for RCP waves occurs, while $A_-$ for LCP waves vanishes, because $n+\gamma=0$. The $A_-$ curve in (a) is identical to the $A_+$ curve in (b).}
\end{figure}

\begin{figure}
\centering\includegraphics[width=9cm]{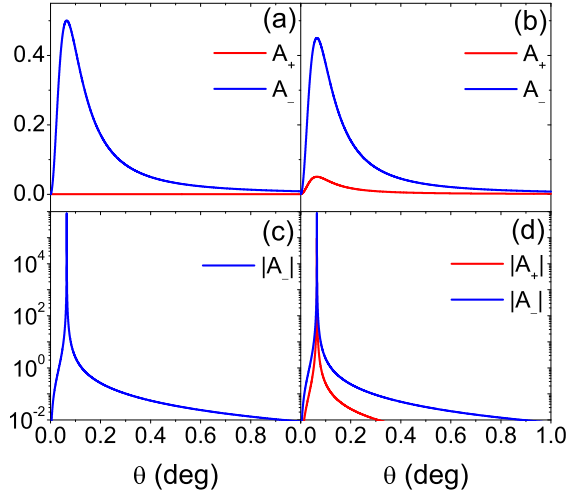}
\caption{Mode conversion coefficients in the absorbing case, $A_+$ and $A_-$, and those in the amplifying case, $\vert A_+\vert$ and $\vert A_-\vert$,
for RCP and LCP waves incident on a uniform chiral slab of thickness $L$ versus incident angle, when (a), (c) $\epsilon_R=\mu_R=\gamma=2$ and (b), (d) $\epsilon_R=4$, $\mu_R=1$, $\gamma=2$. The parameter $k_0 L$ is equal to $5\pi$. Strong absorption occurs in (a) and (b), where $\epsilon_I=\mu_I=10^{-5}$, whereas strong amplification occurs in (c) and (d),
where $\epsilon_I=\mu_I=-10^{-5}$. In (c), $A_+$ is identically zero and is not shown on the logarithmic plot.}
\end{figure}

We first consider the simplest situation in which circularly-polarized waves are incident
on a single uniform slab made of an isotropic chiral medium.
We have verified numerically that in the limit of very small absolute values of the imaginary parts of $\epsilon$ and $\mu$, significant optical absorption or amplification occurs only when $n_+$ vanishes for RCP waves, or when $n_-$ vanishes for LCP waves. If the imaginary parts of $\epsilon$ and $\mu$ are positive, resonant absorption occurs, whereas if they are negative, resonant amplification does.

In Fig.~1, we plot the absorptance, or the mode conversion coefficient, $A_\pm$ for RCP and LCP waves incident on a uniform slab of thickness $L$ with $\epsilon=\epsilon_R+i\epsilon_I$ and $\mu=\mu_R+i\mu_I$ versus the incident angle $\theta$. We consider the four cases where
$\epsilon_R=\mu_R=\gamma=\pm 2$ and $\epsilon_R=\mu_R=-\gamma=\pm 2$, which include both positive and negative index media. The values of $\epsilon_I$ and $\mu_I$ are
$10^{-5}$ and the parameter $k_0L$ is equal to $5\pi$. Strong absorption
for LCP waves occurs at very small incident angles when $n_-=n-\gamma=0$, while $A_+$ for RCP waves vanishes. This corresponds to the cases where $\epsilon_R=\mu_R=\gamma=2$ and $\epsilon_R=\mu_R=\gamma=-2$. In contrast, strong absorption
for RCP waves occurs when $n_+=n+\gamma=0$, while $A_-$ for LCP waves vanishes. This corresponds to the cases where $\epsilon_R=\mu_R=-\gamma=2$ and $\epsilon_R=\mu_R=-\gamma=-2$.
The $A_-$ curve in Fig.~1(a) is identical to the $A_+$ curve in Fig.~1(b). We note that the maximum value of the mode conversion coefficient occurring at $\theta\approx 0.065^\circ$ is very close to 0.5. We have checked numerically that this maximum value remains the same for smaller
values of $\epsilon_I$ and $\mu_I$ and larger values of $k_0L$.

In Fig.~2, we compare the results for the two slabs with $\epsilon_R=\mu_R=\gamma= 2$ and $\epsilon_R=4$, $\mu_R=1$, $\gamma= 2$. In both cases, $n$ is equal to $\gamma$ and strong mode conversion occurs for LCP waves. The main difference between the two is that the impedance $\eta$ in the former case is equal to 1, while that in the latter is equal to 0.5. Because the slabs are surrounded by a vacuum with $\eta=1$, $\eta$ is uniform throughout the space when $\epsilon_R=\mu_R=\gamma= 2$. In this case, RCP and LCP waves are completely decoupled and RCP waves can never cause mode conversion. On the other hand, when $\epsilon_R=4$, $\mu_R=1$, $\gamma= 2$, $\eta$ changes discontinuously at the interfaces of the slab. Incident RCP waves can generate LCP waves at the interfaces, which can subsequently be mode-converted. Therefore strong absorption or amplification occurs for both RCP and LCP waves in this case as shown in Figs.~2(b) and 2(d). Since the coupling between RCP and LCP waves occurs only at the interfaces and is not very strong, the absolute value of $A_+$ is substantially smaller than that of $A_-$. We notice that an extremely large amplification results when inverse mode conversion occurs, as shown in Figs.~2(c) and 2(d).

\begin{figure}
\centering\includegraphics[width=8cm]{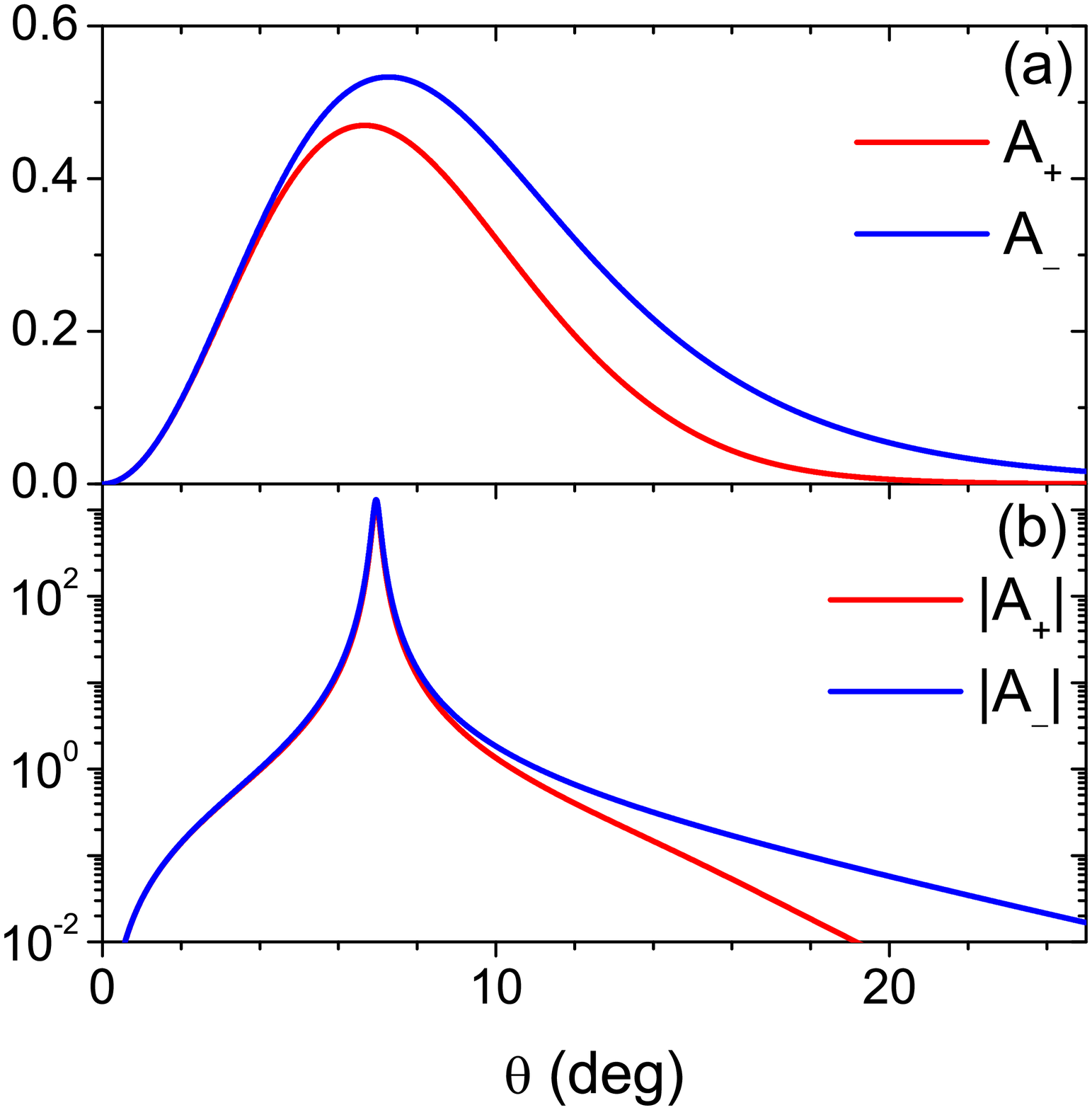}
\caption{Mode conversion coefficients in the absorbing case, $A_+$ and $A_-$, and those in the amplifying case, $\vert A_+\vert$ and $\vert A_-\vert$,
for RCP and LCP waves incident on a nonuniform slab, where $\epsilon_R=\mu_R=2(z/L)-1$ ($0\le z\le L$) and  $\gamma=0.8$, versus incident angle. Circularly-polarized waves are assumed to be incident from the region where $z>L$. The parameter $k_0L$ is equal to $20\pi$. In (a), $\epsilon_I=\mu_I=10^{-8}$ and in (b),
$\epsilon_I=\mu_I=-10^{-8}$.}
\end{figure}

\begin{figure}
\centering\includegraphics[width=8cm]{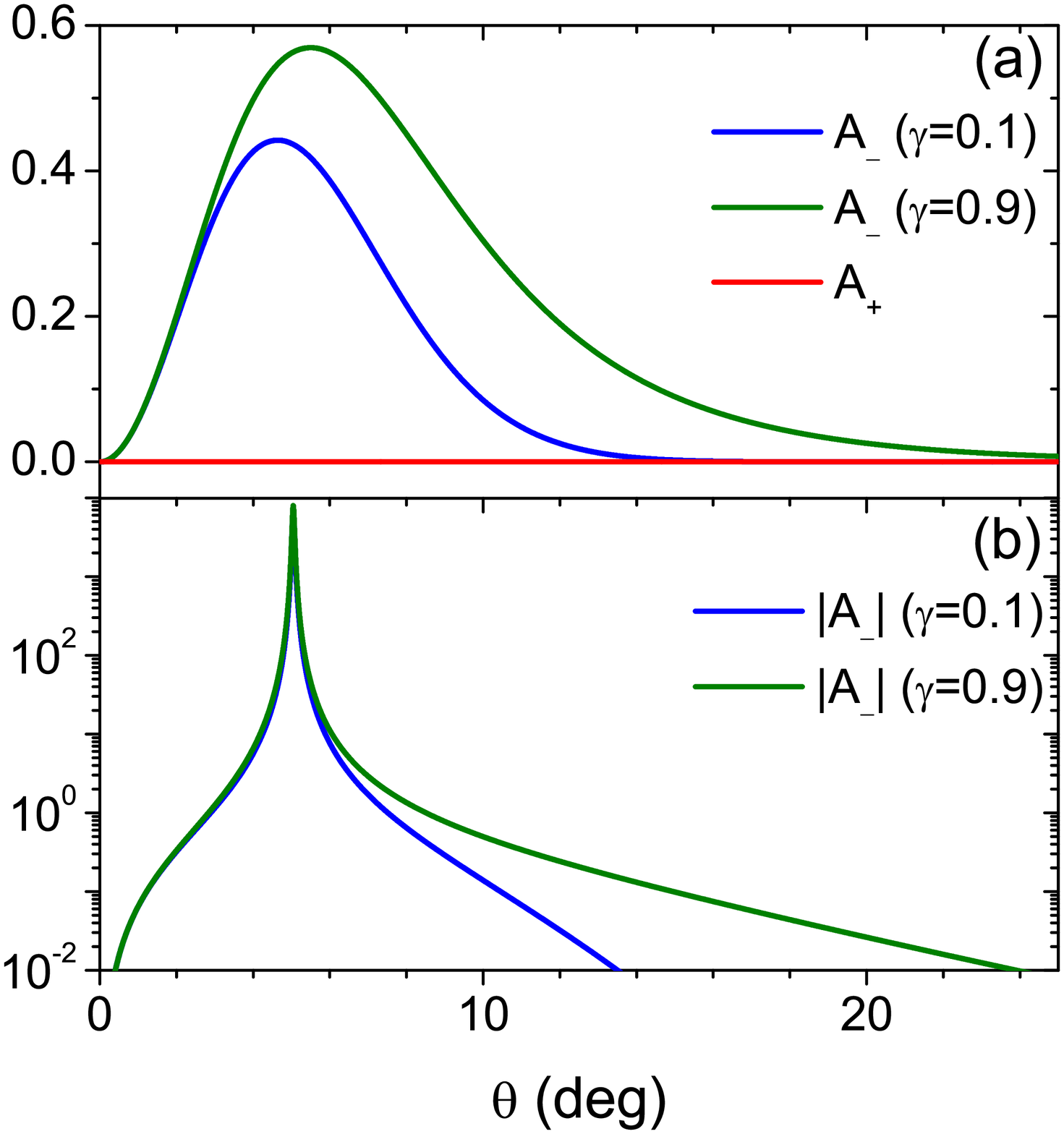}
\caption{Mode conversion coefficients in the absorbing case, $A_+$ and $A_-$, and those in the amplifying case, $\vert A_+\vert$ and $\vert A_-\vert$,
for RCP and LCP waves incident on a nonuniform slab, where $\epsilon_R=\mu_R=z/L$ ($0\le z\le L$) and  $\gamma=0.1$, 0.9, versus incident angle. Circularly-polarized waves are assumed to be incident from the region where $z>L$. The parameter $k_0L$ is equal to $20\pi$. In (a), $\epsilon_I=\mu_I=10^{-8}$ and in (b),
$\epsilon_I=\mu_I=-10^{-8}$. In (b), $A_+$ is identically zero and is not shown on the logarithmic plot.}
\end{figure}

\begin{figure}
\centering\includegraphics[width=8cm]{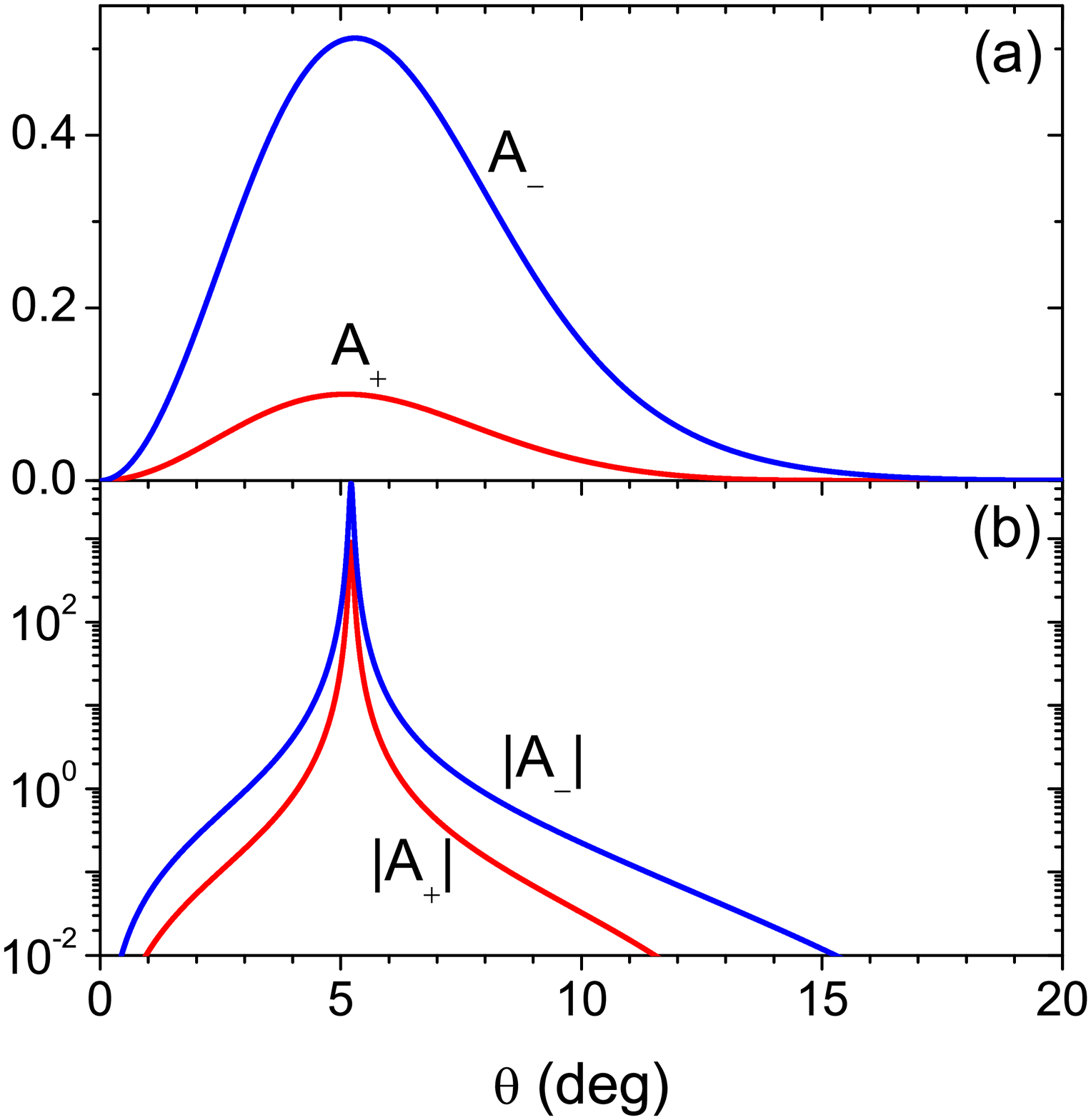}
\caption{Mode conversion coefficients in the absorbing case, $A_+$ and $A_-$, and those in the amplifying case, $\vert A_+\vert$ and $\vert A_-\vert$, for RCP and LCP waves incident on a nonuniform slab, where $\epsilon_R=z/L$, $\mu_R=(z/L)^2$ and $\gamma=0.5$, versus incident angle. Circularly-polarized waves are assumed to be incident from the region where $z>L$. The parameter $k_0L$ is equal to $20\pi$. In (a), $\epsilon_I=\mu_I=10^{-8}$ and in (b),
$\epsilon_I=\mu_I=-10^{-8}$.}
\end{figure}

Next, we consider nonuniform slabs, which contain discrete resonance planes satisfying $n_\pm=0$.
In Fig.~3, we plot the mode conversion coefficients for RCP and LCP waves incident on a nonuniform slab of thickness $L$, where $\epsilon_R=\mu_R=2(z/L)-1$ ($0\le z\le L$) and $\gamma=0.8$, versus incident angle. The parameter $k_0L$ is equal to $20\pi$. In Fig.~3(a), $\epsilon_I$ and $\mu_I$ are equal to $10^{-8}$ and in Fig.~3(b),
they are equal to $-10^{-8}$. The whole curves remain identical for smaller absolute values of $\epsilon_I$ and $\mu_I$, which signifies that the absorption or the amplification is not due to any dissipative damping or gain process, but due to mode conversion or inverse mode conversion. In the present configuration, $n_+$ and $n_-$ vanish at $z/L=(1-\gamma)/2$ and  $z/L=(1+\gamma)/2$ respectively. Therefore, if $\vert\gamma\vert<1$, mode conversion occurs for both RCP and LCP waves. We note that $\eta$ is equal to 1 everywhere, and therefore RCP and LCP waves are independent. When $\gamma$ is 0.8, $n_+$ and $n_-$ vanish at $z=0.1L$ and $z=0.9L$ respectively. LCP waves incident from the region where $z>L$ arrive at $z=0.9L$ and are mode-converted there, while RCP waves have to propagate for much longer distances to $z=0.1L$ before getting mode-converted. The increased reflection from the inhomogeneity experienced by RCP waves makes $\vert A_+\vert$ smaller than $\vert A_-\vert$, especially at larger incident angles.

In Fig.~4, we consider a nonuniform slab with a different configuration such that $\epsilon_R=\mu_R=z/L$ ($0\le z\le L$). In this case, $n_+$ never vanishes, while $n_-=0$ at $z/L=\gamma$. Because the impedance is uniform and equal to 1 everywhere, RCP waves are unable to cause mode conversion and $A_+$ is identically zero. We compare the results for the two cases where $\gamma=0.1$ and $\gamma=0.9$. Resonant absorption and amplification are substantially stronger for $\gamma=0.9$, especially at larger incident angles, than for $\gamma=0.1$, which is due
to the fact that waves have to
propagate for longer distances to reach the resonance plane experiencing further reflections in the case of $\gamma=0.1$.

Finally, in Fig.~5, we consider the configuration where $\epsilon_R=z/L$ and $\mu_R=(z/L)^2$ ($0\le z\le L$). Then the real parts of the refractive index and the impedance are given by $(z/L)^{3/2}$ and $(z/L)^{1/2}$ respectively. Because the impedance is nonuniform, RCP and LCP waves are coupled to each other. Similarly to the previous case, $n_+$ never vanishes and $n_-=0$ at $z/L=\gamma^{2/3}$. The inhomogeneity of $\eta$ causes a strong coupling between RCP and LCP waves, which results in a substantial amount of $\vert A_+\vert$, as shown in Fig.~5.

\section{Conclusion}
\label{sec4}

In this paper, we have studied the mode conversion and the resonant absorption and amplification of circularly-polarized electromagnetic waves in stratified chiral media theoretically using a generalized version of the invariant imbedding method. In uniform isotropic chiral media, RCP and LCP waves are the eigenmodes with different effective refractive indices $n_+$ and $n_-$, whereas in the chiral media with a nonuniform impedance variation, they are not the eigenmodes and are coupled to each other. We have found that both in uniform chiral slabs where $n_+$ or $n_-$ is near zero and in chiral transition metamaterials where $n_+$, $n_-$ or both change from positive to negative values, a strong absorption or amplification of circularly-polarized waves occurs in the presence of a vanishingly small amount of damping or gain. We have presented numerical calculations of the mode conversion coefficient for various spatial configurations of $\epsilon$ and $\mu$ with an emphasis on the influence of a nonuniform impedance. Based on the investigations reported here, it is possible to develop an efficient absorber or amplifier that reacts selectively to the helicity of the circular polarization. The strong enhancement of the electromagnetic fields can be exploited in developing novel nonlinear optical devices operating with circularly-polarized waves.

\section*{Acknowledgments}
This work has been supported by the National Research Foundation of Korea Grants (NRF-2012R1A1A2044201, NRF-2015R1A2A2A01003494) funded by the Korean Government.


\begin{thebibliography}{99}

\bibitem{1} N. Garcia, E. V. Ponizovskaya, and John Q. Xiao, ``Zero permittivity materials: Band gaps at the visible,'' Appl. Phys. Lett. {\bf 80}, 1120--1122 (2002).

\bibitem{2} M. Silveirinha and N. Engheta, ``Tunneling of electromagnetic energy through subwavelength channels and bends using $\epsilon$-near-zero materials,'' Phys. Rev. Lett. {\bf 97}, 157403 (2006).

\bibitem{3} R. Liu, Q. Cheng, T. Hand, J. J. Mock, T. J. Cui, S. A. Cummer, and D. R. Smith, ``Experimental demonstration of electromagnetic tunneling through an epsilon-near-zero metamaterial at microwave frequencies,'' Phys. Rev. Lett. {\bf 100}, 023903 (2008).

\bibitem{4} A. Ciattoni, C. Rizza, and E. Palange, ``Extreme nonlinear electrodynamics in metamaterials with very small linear dielectric permittivity,'' Phys. Rev. A {\bf 81}, 043839 (2010).

\bibitem{5} Y. Jin, S. Xiao, N. A. Mortensen, and S. He, ``Arbitrarily thin metamaterial structure for perfect absorption and giant magnification,'' Opt. Express {\bf 19}, 11114--11119 (2011).

\bibitem{6} M. A. Vincenti, D. de Ceglia, A. Ciattoni, and M. Scalora, ``Singularity-driven second-and third-harmonic generation at $\epsilon$-near-zero crossing points,'' Phys. Rev. A {\bf 84}, 063826 (2011).

\bibitem{7} S. Feng and K. Halterman, ``Coherent perfect absorption in epsilon-near-zero metamaterials,'' Phys. Rev. B {\bf 86}, 165103 (2012).

\bibitem{8} S. Vassant, A. Archambault, F. Marquier, F. Pardo, U. Gennser, A. Cavanna, J. L. Pelouard, and J. J. Greffet, ``Epsilon-near-zero mode for active optoelectronic devices,'' Phys. Rev. Lett. {\bf 109}, 237401 (2012).

\bibitem{9} S. Campione, D. de Ceglia, M. A. Vincenti, M. Scalora, and F. Capolino, ``Electric field enhancement in $\epsilon$-near-zero slabs under TM-polarized oblique incidence,'' Phys. Rev. B {\bf 87}, 035120 (2013).

\bibitem{10} S. Zhong and S. He, ``Ultrathin and lightweight microwave absorbers made of mu-near-zero metamaterials,'' Sci. Rep. {\bf 3}, 2083 (2013).

\bibitem{11} P. Ginzburg, F. J. Rodr\'{i}guez Fortu\~{n}o, G. A. Wurtz, W. Dickson, A. Murphy, F. Morgan, R. J. Pollard, I. Iorsh, A. Atrashchenko, P. A. Belov, Y. S. Kivshar, A. Nevet, G. Ankonina, M. Orenstein, and A. V. Zayats, ``Manipulating polarization of light with ultrathin epsilon-near-zero metamaterials,'' Opt. Express {\bf 21}, 14907--14917 (2013).

\bibitem{12} D. de Ceglia, S. Campione, M. A. Vincenti, F. Capolino, and M. Scalora, ``Low-damping epsilon-near-zero slabs: Nonlinear and nonlocal optical properties,'' Phys. Rev. B {\bf 87}, 155140 (2013).

\bibitem{13} K. Halterman and J. M. Elson, ``Near-perfect absorption in epsilon-near-zero structures with hyperbolic dispersion,'' Opt. Express {\bf 22}, 7337--7348 (2014).

\bibitem{14} J. Yoon, M. Zhou, M. A. Badsha, T. Y. Kim, Y. C. Jun, and C. K. Hwangbo, ``Broadband epsilon-near-zero perfect absorption in the near-infrared,'' Sci. Rep. {\bf 5}, 12788 (2015).

\bibitem{15} K. Kim, D.-H. Lee, and H. Lim, ``Resonant absorption and mode conversion in a transition layer between positive-index and negative-index media,'' Opt. Express {\bf 16}, 18505--18513 (2008).

\bibitem{16} N. M. Litchinitser, A. I. Maimistov, I. R. Gabitov, R. Z. Sagdeev, and
V. M. Shalaev, ``Metamaterials: electromagnetic enhancement at zero-index transition,'' Opt. Lett. {\bf 33}, 2350--2352 (2008).

\bibitem{17} I. Mozjerin, E. A. Gibson, E. P. Furlani, I. R. Gabitov, and N. M. Litchinitser, ``Electromagnetic enhancement in lossy optical transition metamaterials,'' Opt. Lett. {\bf 35}, 3240--3242 (2010).

\bibitem{18} Y. S. Ding, C. T. Chan, and R. P. Wang, ``Optical waves in a gradient negative-index lens of a half-infinite length,'' Sci. Rep. {\bf 3}, 2954 (2013).

\bibitem{19} I. D. Rukhlenko, ``Optical propagation through graded-index metamaterials in the presence of gain,'' Plasmonics {\bf 9}, 1257--1263 (2014).

\bibitem{20} Z. A. Kudyshev, I. R Gabitov, A. I. Maimistov, R. Z Sagdeev, and N. M. Litchinitser, ``Second harmonic generation in transition metamaterials,'' J. Opt. {\bf 16}, 114011 (2014).

\bibitem{21} J. Sun, X. Liu, J. Zhou, Z. Kudyshev, and N. M. Litchinitser, ``Experimental demonstration of anomalous field enhancement in all-dielectric transition magnetic metamaterials,'' Sci. Rep. {\bf 5}, 16154 (2015).









\bibitem{22} D. E. Hinkel-Lipsker, B. D. Fried, and G. J.
Morales, ``Analytic expressions for mode conversion in a plasma with
a linear density profile,'' Phys. Fluids B {\bf 4}, 559--575 (1992).

\bibitem{23}
K. Kim and D.-H. Lee, ``Invariant imbedding theory of mode
conversion in inhomogeneous plasmas. I. Exact calculation of the
mode conversion coefficient in cold, unmagnetized plasmas,'' Phys.
Plasmas {\bf 12}, 062101 (2005).

\bibitem{24} D. J. Yu, K. Kim, and D.-H. Lee, ``Resonant enhancement of mode conversion in unmagnetized plasmas due to a periodic density modulation superimposed on a linear electron density profile,'' Phys. Plasmas {\bf 17}, 102110 (2010).

\bibitem{25} D. J. Yu, K. Kim, and D.-H. Lee, ``Temperature dependence of mode conversion in warm, unmagnetized plasmas with a linear density profile,'' Phys. Plasmas {\bf 20}, 062109 (2013).

\bibitem{26} D. J. Yu and K. Kim, ``Effects of a random spatial variation of the plasma density on the mode conversion in cold, unmagnetized, and stratified plasmas,'' Phys. Plasmas {\bf 20}, 122104 (2013).

\bibitem{27} V. L. Ginzburg, {\it The Propagation of Electromagnetic Waves in Plasmas}
(Pergamon, 1970).

\bibitem{28} K. G. Budden, {\it The Propagation of Radio Waves} (Cambridge University
Press, 1985).

\bibitem{29} D. G. Swanson,
{\it Theory of Mode Conversion and Tunneling in Inhomogeneous
Plasmas} (Wiley, 1998).

\bibitem{30} E. Mj{\o}lhus, ``On linear conversion in a magnetized plasma,'' Radio Sci. {\bf 25}, 1321--1339 (1990).

\bibitem{31} K. Kim and D.-H. Lee, ``Invariant imbedding theory of mode conversion in inhomogeneous
plasmas. II. Mode conversion in cold, magnetized plasmas with
perpendicular inhomogeneity,'' Phys. Plasmas {\bf 13}, 042103
(2006).

\bibitem{32} E.-H. Kim, I. H. Cairns, and P. A. Robinson, ``Extraordinary-mode radiation produced by linear-mode conversion of Langmuir waves,'' Phys. Rev. Lett. {\bf 99}, 015003 (2007).

\bibitem{33} I. V. Lindell, A. H. Sihvola, S. A. Tretyakov, and A. J. Viitanen, {\it Electromagnetic Waves
in Chiral and Bi-Isotropic Media} (Artech House, 1994).

\bibitem{34} J. Lekner, ``Optical properties of isotropic chiral media,'' Pure Appl. Opt.
{\bf 5}, 417--443 (1996).

\bibitem{35} J. B. Pendry, ``A chiral route to negative refraction,''
Science {\bf 306}, 1353--1355 (2004).

\bibitem{36} K. Kim, H. Yoo, and H. Lim, ``Exact analytical expressions for the dispersion relation
of one-dimensional chiral photonic crystals,'' Waves Random Complex
Media {\bf 16}, 75--84 (2006).

\bibitem{37} Y. Tamayama, T. Nakanishi, K. Sugiyama, and M. Kitano, ``An invisible medium for circularly polarized electromagnetic waves,'' Opt. Express {\bf 16}, 20869--20875 (2008).

\bibitem{38} B. Wang, J. Zhou, T. Koschny,
M. Kafesaki, and C. M. Soukoulis, ``Chiral metamaterials: simulations and
experiments,'' J. Opt. A: Pure Appl. Opt. {\bf 11}, 114003 (2009).

\bibitem{39} E. Plum, J. Zhou, J. Dong, V. A. Fedotov, T. Koschny, C. M. Soukoulis, and N. I. Zheludev,
``Metamaterial with negative index due to chirality,'' Phys. Rev. B {\bf 79}, 035407 (2009).

\bibitem{40} S. Zhang, Y.-S. Park, J. Li, X. Lu, W. Zhang, and X. Zhang, ``Negative refractive index in chiral metamaterials,'' Phys. Rev. Lett. {\bf 102}, 023901 (2009).

\bibitem{41} Y. Y. Huang, W. T. Dong, L. Gao, and C. W. Qiu, ``Large positive and negative lateral shifts near pseudo-Brewster dip on reflection from a chiral metamaterial slab,'' Opt. Express {\bf 19}, 1310--1323 (2011).

\bibitem{42} Z. Li, M. Mutlu, and E. Ozbay, ``Chiral metamaterials: from optical
activity and negative refractive index to
asymmetric transmission,'' J. Opt. {\bf 15}, 023001 (2013).

\bibitem{43} K. J. Lee, J. W. Wu, and K. Kim, ``Defect modes in a one-dimensional photonic crystal with a chiral defect layer,'' Opt. Mater. Express {\bf 4}, 2542--2550 (2014).

\bibitem{44} Y. Cao and J. Li, ``Complete band gaps in one-dimensional photonic crystals with negative refraction arising from strong chirality,'' Phys. Rev. B {\bf 89}, 115420 (2014).


\bibitem{45}
V. I. Klyatskin, ``The imbedding method in statistical
boundary-value wave problems,'' Prog. Opt. {\bf 33}, 1--127 (1994).

\bibitem{46} K. Kim, H. Lim, and D.-H. Lee, ``Invariant imbedding
equations for electromagnetic waves in stratified magnetic media:
Applications to one-dimensional photonic crystals,'' J. Korean Phys.
Soc. {\bf 39}, L956--L960 (2001).

\bibitem{47} K. Kim, D.-H. Lee, and H. Lim, ``Theory of the propagation of coupled waves in
arbitrarily inhomogeneous stratified media,'' EPL {\bf
69}, 207--213 (2005).

\bibitem{48}
K. Kim, D. K. Phung, F. Rotermund, and H. Lim, ``Propagation of
electromagnetic waves in stratified media with nonlinearity in both
dielectric and magnetic responses,'' Opt. Express {\bf 16},
1150--1164 (2008).


\end{thebibliography}
\end{document}